# All-wurtzite (In,Ga)As-(Ga,Mn)As core-shell nanowires grown by molecular beam epitaxy

*Aloyzas Šiušys[†,\*], Janusz Sadowski[‡,†,\*\*], Maciej Sawicki[†], Sławomir Kret[†], Tomasz Wojciechowski[†], Katarzyna Gas[†], Wojciech Szuszkiewicz[†], Agnieszka Kaminska[§] and Tomasz Story[†]*

[†] Institute of Physics, Polish Academy of Sciences, al. Lotników 32/46, PL 02-668 Warszawa, Poland

[‡] MAX-IV Laboratory, Lund University, P.O. Box 118, SE 221 00 Lund, Sweden

[§] Institute of Physical Chemistry, Polish Academy of Sciences, Kasprzaka 44/52, PL 01-224 Warszawa, Poland

ABSTRACT

Structural and magnetic properties of (In,Ga)As-(Ga,Mn)As core-shell nanowires grown by molecular beam epitaxy on GaAs(111)B substrate with gold catalyst have been investigated.

To whom correspondence should be addressed:
[\*] siusys@ifpan.edu.pl
[\*\*] janusz.sadowski@maxlab.lu.se

(In,Ga)As core nanowires were grown at high temperature (500 °C) whereas (Ga,Mn)As shells were deposited on the {1-100} side facets of the cores at much lower temperature (220 °C). High resolution transmission electron microscopy images and high spectral resolution Raman scattering data show that both the cores and the shells of the nanowires have wurtzite crystalline structure. Scanning and transmission electron microscopy observations show smooth (Ga,Mn)As shells containing 5% of Mn epitaxially deposited on (In,Ga)As cores containing about 10% of In, without any misfit dislocations at the core-shell interface. With the In content in the (In,Ga)As cores larger than 5% the (In,Ga)As lattice parameter is higher than that of (Ga,Mn)As and the shell is in the tensile strain state. Elaborated magnetic studies indicate the presence of ferromagnetic coupling in (Ga,Mn)As shells at the temperatures in excess of 33 K. This coupling is maintained only in separated mesoscopic volumes resulting in an overall superparamagnetic behavior which gets blocked below ~17 K.

(Ga,Mn)As is a canonical (III,Mn)-V dilute ferromagnetic semiconductor (FMS) combining semiconducting and magnetic properties in one material.[1] Thin layers of (Ga,Mn)As[2] as well as other (III,Mn)-V FMS such as (In,Mn)As,[3] (Ga,Mn)Sb[4] and (In,Mn)Sb[5] have been extensively investigated over the last two decades. The wide range of new physical phenomena and functionalities have been discovered in FMS due to the possibility of tuning their magnetic properties by the methods routinely used for modifying the electronic properties of semiconductors, such as application of electric fields by electrostatic gates,[6,7,8,9] pressure[10] or irradiation with light.[11,12,13] In this context using quasi one-dimensional (1D) geometry of nanowires is advantageous since it enhances the possibility of controlling the electronic properties up to the ultimate level of the single carrier.[14,15,16]

Since (Ga,Mn)As layers with Mn content exceeding 1%, which is essential for the ferromagnetic phase transition to occur,[1] can only be grown at low temperatures, they contain substantial amount of defects.[17,18,19] Hence it is interesting to exploit different growth methods in order to investigate various possibilities of obtaining material with optimized properties. One of the new technological routes is to employ a 1D nanowire growth mode which differs substantially from the two-dimensional (2D) MBE growth of the layers. In 1D structure it is possible to efficiently accommodate stress of lattice mismatched materials in radial/axial heterostructures. Furthermore crystalline structures different than naturally occurring in 3D (bulk) and 2D (layer) can arise in the 1D (nanowire) case. Core-shell nanowires (NWs) reported here crystallize in the hexagonal (wurtzite) phase [both (In,Ga)As cores and (Ga,Mn)As shells], whereas in a 2D (and 3D) case both materials can only be crystallized in the cubic (zinc-blende) structure. Thus growth of the 1D (nanowire) structures makes it possible to fabricate heterostructures of a new type, which are impossible to obtain in the planar geometry.

Ferromagnetic nanowires are extensively investigated as building blocks of future memory devices. An important role is played, in particular, by the magnetic domain walls which can be moved along the nanowires.[20] Manufacturing (Ga,Mn)As in the form of nanowires by a bottom-up self-organizing approach, avoiding defects induced when preparing nanowires lithographically, would help to control their magneto-electronic properties better.

To this end, we have experimentally tested various MBE growth methods of incorporation of Mn magnetic ions into GaAs in the nanowire geometry. GaAs NWs grow best at high temperatures (550 °C - 650 °C), thus first we have investigated the Mn doping limits of GaAs NWs grown in this high temperature range.[21] We have observed that Mn doesn't deteriorate the growth of GaAs NWs, even with quite high Mn/Ga flux ratio of about 3%. We have found that at high growth temperatures, Mn can be introduced into GaAs NWs only at doping levels (below $10^{18}$ cm$^{-3}$).[21] Such nanowires grow well in <111> preferential crystal directions, revealing hexagonal cross-section with a catalyst droplet on the top.[21] It is well known that Mn solubility limits in the GaAs lattice can be overcome at non equilibrium growth conditions using low growth temperature (LT-MBE).[1] Therefore, previously we have also investigated the growth of NWs at low temperatures (about 350 °C) only slightly higher than those typically applied for MBE growth of (Ga,Mn)As layers. However, low temperatures are detrimental to the growth of NWs. The LT-grown NWs are irregular, tapered, branched and have low crystalline quality.[22] Although the concentration of Mn in the LT grown (Ga,Mn)As NWs is larger than in the NWs grown at high temperature, it is still well below 1% which is too low to support the ferromagnetic state. Moreover the presence of segregated MnAs renders the analysis of magnetic properties of these NWs difficult.

In order to achieve high Mn concentration, the NWs can be grown by MBE by a two-stage method in the form of radial core-shell heterostructures, with primary III-As cores grown at high temperature followed by (Ga,Mn)As shells grown at low temperature. Few attempts to prepare such NWs with pure zinc-blende GaAs cores have been reported previously.[23,24] Interestingly, an enhanced magnetic response at temperatures below 20 K was reported there, which served as a stimulus for further research, augmented by recent theoretical results which predicted an increase of $T_c$ for (Ga,Mn)As NWs crystallizing in wurtzite structure.[25,26,27]

In this work we investigate the all-wurtzite (In,Ga)As-(Ga,Mn)As core-shell NWs. To our knowledge, experimental results concerning such structures have not been reported yet.

Nanowires exploited here have been grown by the MBE method employing vapor-liquid-solid (VLS) growth mode. Prior to the growth of the nanowires a small amount of gold (equivalent of a 2 Å thick continuous layer) was deposited on epi-ready GaAs(111)B substrate at low temperature (100 °C) in the MBE system dedicated for metals. Then the substrate was transferred (in the air) to the III-V semiconductor MBE system and annealed at 590 °C to desorb the native oxide from GaAs substrate and to create small nanometer-scale Au catalysts on it. Core-shell NWs were grown in two steps. In the first step, (In,Ga)As primary cores were grown at high temperature (growth temperature about 500 °C). The NWs are oriented vertically with respect to the substrate surface; they are up to 1.6 micrometer long and have diameters of 70-100 nm. The nominal In content in the core was chosen to be around 13%. It has been calibrated by the standard procedures based on RHEED intensity oscillations taken during growth of GaAs and InGaAs layers on a test GaAs(100) substrate without gold catalyst. The actual concentration of In in the (In,Ga)As NWs is most likely lower i.e. equals about 10% due to the different growth mechanism involved in the MBE growth of planar (In,Ga)As layers as compared to the Au-catalyst induced growth of (In,Ga)As NWs at identical conditions, in agreement with previous reports (e.g. about 20% lower incorporation of In into wurtzite (In,Ga)As NWs compared to the In content in the (In,Ga)As(100) planar layer has been reported by Jabeen et. al.[28]).

The growth rate of the NWs has been observed to be almost 5 times faster than that of the layers grown with the same III and V element fluxes, and is equal to about 1 µm/h. Density of NWs is high (about $2 \times 10^9$ NWs/cm$^2$) as indicated in the scanning electron microscopy (SEM) images displayed in Fig. 1.

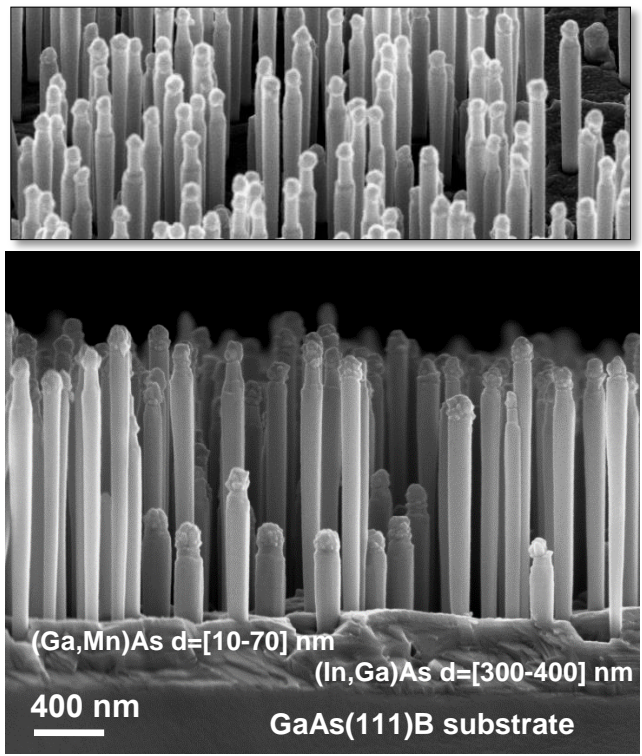

**Figure 1.** SEM images of Au-catalyzed (In,Ga)As-(Ga,Mn)As nanowires grown on GaAs(111)B substrate by MBE. (In,Ga)As and (Ga,Mn)As layers deposited in-between the nanowires can be identified in the vicinity of the substrate surface. Birds-eye view of nanowires (from the center part of the sample) is shown in the top panel.

After the (In,Ga)As NW cores growth, the substrate temperature was reduced to around 220 °C and the low temperature (Ga,Mn)As shells were deposited. The quality of the (Ga,Mn)As shells is strongly dependent on the growth temperature and Mn content. Smooth shells can only be obtained in a narrow substrate temperature window (220°C - 240°C). The growth conditions of the shells correspond to intended Mn content of 5%. The shells' growth rate is about 20 nm/h. This is much lower than the planar growth rate of 200 nm/h, for the (Ga,Mn)As layer growth with the same group III and group V element fluxes. Since parts of the substrate not covered by NWs are exposed to the fluxes during the MBE growth of the shells, the (Ga,Mn)As layer is also deposited in between the NWs. It should also be noted that during the prevailing growth of (In,Ga)As nanowires a 300 nm - 400 nm thick layer of (In,Ga)As in between the NWs grows too. Analysis of the high-contrast SEM images shows that the thickness of the (Ga,Mn)As layer deposited in-between the NWs can be up to 70 nm depending on the alternating level of shadowing effect due to the inhomogeneous distribution of NWs (the thickness of the planar (Ga,Mn)As layer grown on a GaAs(111)B substrate without catalytic gold nanoparticles would be 100 nm). This has important consequences for the measurements of magnetic properties of such samples as explained further in the text.

The structural and chemical characterization of the samples was carried out by transmission electron microscopy (TEM) methods. The growth direction of the NWs and their morphology was studied in FEI Titan 80-300 field-emission gun (TEM) operated at 300 kV with aberration correction of objective lens, which enables a high resolution TEM (HRTEM) investigation of the crystal structure with near atomic resolution. The first sample for TEM studies was prepared in the following way: NWs were chopped from the substrate and suspended on a carbon grid. Figure 2 shows the TEM image of a short section of single (In,Ga)As-(Ga,Mn)As core-shell NW with <0001> orientation of its axis. It is easy to distinguish between the (In,Ga)As core and 12 nanometer-thick (Ga,Mn)As shell. Moreover, the crystalline structure of the shell is perfect. There are no misfit dislocations and the shell is in epitaxial relation to the core with sharp and coherent interface between the core and the shell. In contrast, the shell covering gold catalyst is polycrystalline and of irregular shape (not shown). About 3 nm thick native oxide layer on the outer surface of the shell is observed after 10 s of Ar-O ions pre-cleaning before loading the specimen into the TEM chamber.

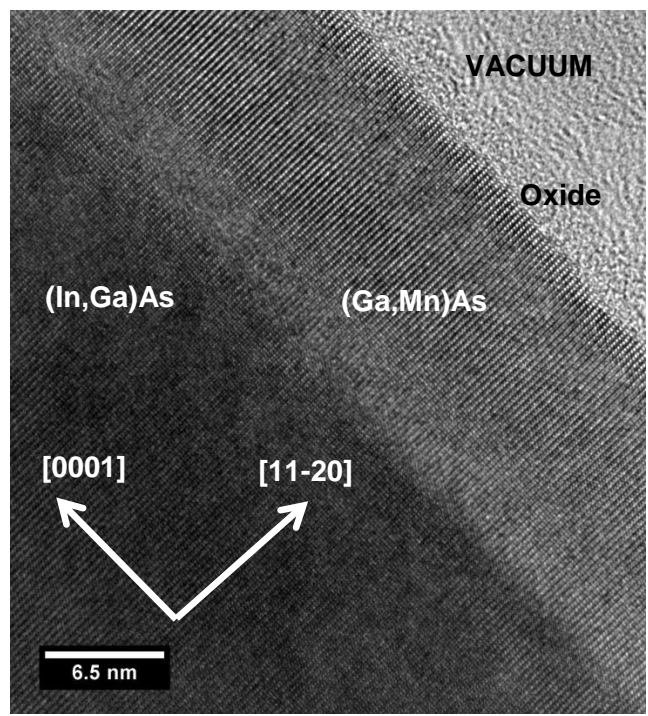

**Figure 2.** High resolution TEM image of (In,Ga)As-(Ga,Mn)As core-shell NW interface. The (Ga,Mn)As is epitaxial with respect to the (In,Ga)As core without any dislocations at the interface.

The chemical composition of individual NWs was investigated by high-angle annular dark field scanning transmission electron microscopy (HAADF-STEM) method combined with the energy-dispersive X-ray spectroscopy (EDS). The scanning TEM (STEM) images of the whole NW and the top part of a single (In,Ga)As-(Ga,Mn)As core-shell NW are shown in Figs. 3a and 3b, respectively. All components of the core-shell NW: the shell, the core (both of the wurtzite crystalline structure) and the polycrystalline gold droplet with irregular shell at the top of NW can be identified. The gold catalyst is visible as the bright area at the top of the NW core. The shell material deposited on the gold catalyst is polycrystalline, thus it has the irregular shape. For several NWs, basal stacking faults were found in their top parts. The EDS analysis of the (In,Ga)As-(Ga,Mn)As core-shell NW confirmed the presence of Mn in the shell and Mn content of 5% ($\pm$ 0.5%). No local precipitations of the Mn, within the EDS detection limit of 0.5 %, were found.

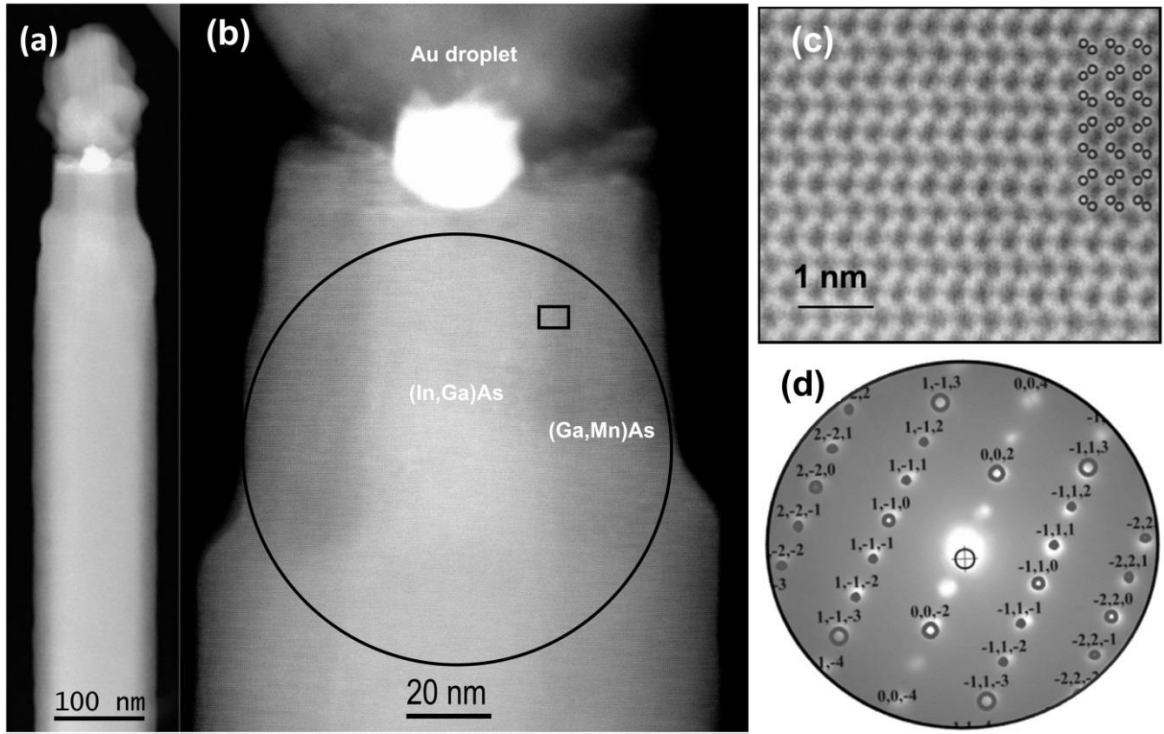

**Figure 3.** (a) STEM image of (In,Ga)As-(Ga,Mn)As NW. The (In,Ga)As core can be distinguished as slightly brighter than the (Ga,Mn)As shell. (b) HR-STEM image of the top part of (Ga,Mn)As-(In,Ga)As NW; (c) atomic positions in the small section of NW marked by the rectangle in panel (b), the right part shows the theoretical atomic positions in the wurtzite structure; (d) – indexed Bragg reflections of the TEM selected area electron diffraction image in the [11-20] zone axis taken from area marked by the circle in panel (b).

High resolution STEM HAADF image which is shown in Fig.3, clearly shows the atomic order characteristic for wurzite crystal projected in [11-20] direction (see zoomed part in Fig. 3c). Also in the case of selected area electron diffraction obtained for larger part of the NW, which is shown by the circle in Fig.3b (the diameter of used slit is 100 nm) the theoretical kinematical pattern for [11-20] zone axis fits well to the experimental image (see Fig. 3d). This confirms the presence of a single phase wurtzite structure for both the (In,Ga)As core and the (Ga,Mn)As shell.

TEM investigations were performed in a large range of magnification from 100x to 460kx to investigate the crystal perfection of ensemble of NWs. More than 100 NWs were checked for the presence of structural imperfections such as basal SF or cubic segments. We can conclude that there is more than 15% of NW with low SF density (less than 5 SF which appear mainly at the ends of NWs). Fig. 4 shows a representative NW with low defect density viewed by HRTEM and HR-STEM in [11-20] zone axis. The image of the whole NW is shown in Fig. 4a. The frames 1, 2 and 3 indicate the area zoomed in panels (b), (d) and (e), respectively. The whole middle part of the NW has perfect WZ structure (see fig 4d) without any structural defects. Few SFs can be identified in the bottom part of the NW (frame 1 in Fig. 4a). Two of them (indicated by arrows, separated by 76 nm) are well visible in Fig. 4b. Remaining four SFs (not shown) at the end of the NW are separated by 46-80 nm. The atomic structure of individual SF is shown on the HR-STEM image in Fig 4c. Two stacking faults can also be seen in the top section of NW, about 30 nm from the gold droplet (see Fig 4f).

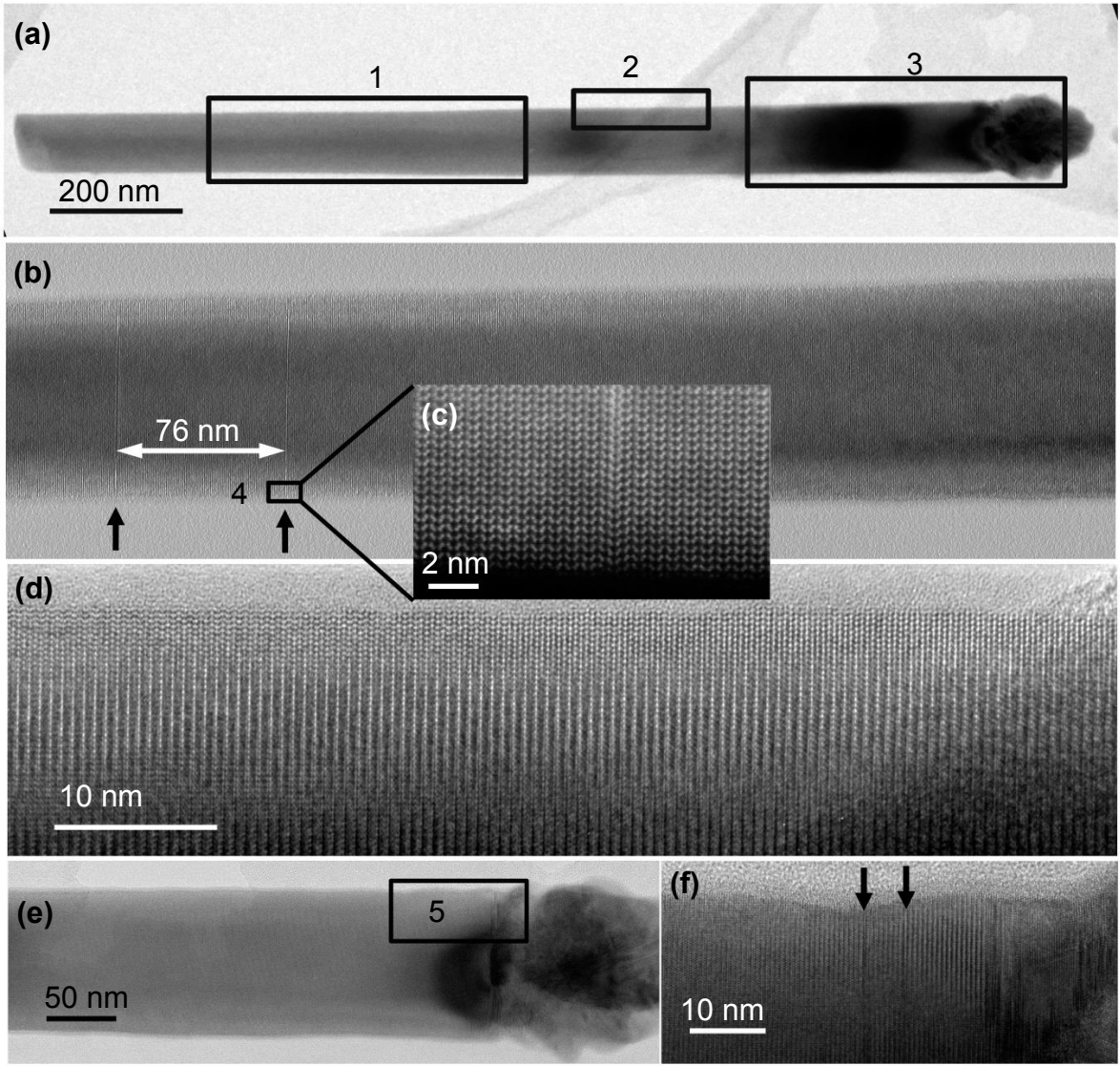

**Figure 4.** (a) TEM image of almost defect free wurtzite NW in [11-20] zone axis; (b) zoomed part of the NW section selected in frame 1; (c) HR-STEM image of the frame 4 with one SF; (d) HRTEM image of the defect-free NW part (frame 2), (e) top part of the NW confined in the frame 3; (f) zoomed HRTEM image of the topmost part of the NW (frame 5), arrows indicate the presence of SF.

(In,Ga)As-(Ga,Mn)As NWs have also been investigated by TEM in the cross-sectional geometry (with cross-sections perpendicular to the NW axis), following the preparation method described below. First, NWs were transferred onto the Si(100) substrate covered with native SiO$_2$ layer.

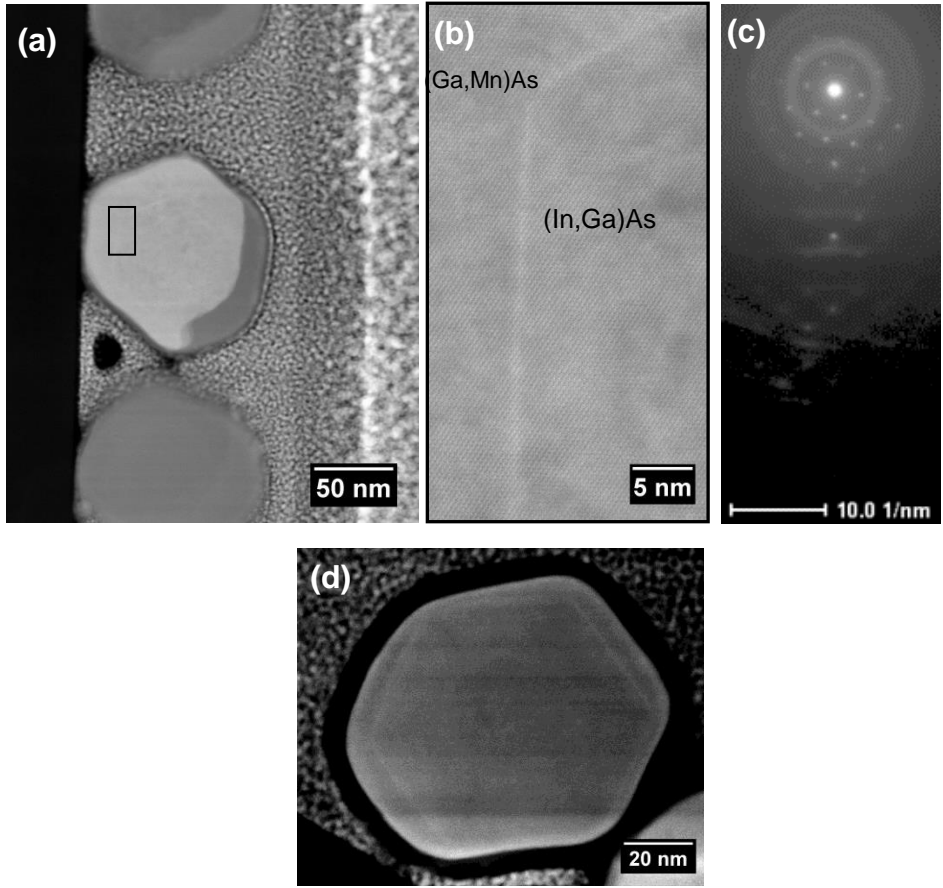

**Figure 5.** (a) - STEM image of a thin FIB lamella containing three NWs embedded in the platinum. The NWs have either hexagonal or round cross-sections depending on the thickness of the shell; (b) –zoomed part of the high resolution STEM image from panel (a) in the frame showing (In,Ga)As core and epitaxial (Ga,Mn)As shell; (c) selected area electron diffraction patterns containing contribution from different parts of the TEM specimen: rings - polycrystalline platinum cover; upper spot diffraction pattern - NW; lower diffraction pattern - Si substrate; (d) - cross-section from the upper part of the single NW with the shell of non-homogenous thickness.

After selection of suitable NWs and marking their position, substrate and NWs were capped with 1.2 μm thick platinum layer. A TEM specimen was cut out using focused gallium ion beam (FIB) technique. An example of successful cut containing three NWs is shown in the STEM image in Fig.5a. The NW located in the middle is perfectly aligned to [0001] zone axis. Part of this NW appearing as darker is amorphous, as confirmed by high magnification HRTEM images (not shown). This partial amorphization of the tops of NWs from one side is due to high electron beam intensity applied during the platinum deposition process. This kind of amorphization didn't occur for the NW shown in Fig 5d due to the lower e-beam intensity used for platinum deposition in this case.

Investigated NWs have either hexagonal or round cross-sections depending on the shell thickness and on spatial arrangement of nanowires causing inhomogeneous shadowing. The high temperature grown (In,Ga)As NW cores have perfect hexagonal cross-sections. Due to the enhanced shadowing of bottom parts of NWs during the shell deposition the (Ga,Mn)A shells thickness is increasing slightly along the NW lengths (smaller at the close-to-the substrate parts, larger at the top part). By this it is possible to determine if given cross-section originates from the top, middle or bottom part of NW. The NW shown in Fig. 5d has small shell thickness reaching 7 nm in the thickest place. Figure 5b shows a zoomed image of a small part of the NW cross-section (selected in the rectangular frame displayed in Fig. 5a). The interface between (Ga,Mn)As shell and (In,Ga)As core is sharp, and contains no structural defects. Selected area diffraction patterns have been collected from one of the NW cross-sections. The ring patterns of diffracted electrons originate from platinum; spotty patterns from NW and Si substrate (see Fig.5 c).

To further confirm the wurtzite structure, the chemical composition, as well as to estimate the carrier concentration in the NW shells we have performed room temperature Raman scattering (RS) measurements for the NWs transferred from the growth substrate onto a Si(100) wafer by ultra-sonification. The measurements have been performed for statistically relevant number of NWs.

The room temperature RS measurements have been performed using a Mono Vista CRS Confocal Laser Raman System with a spectral resolution close to 0.35 cm$^{-1}$. The spectra have been collected using 532 nm laser light. The measurements have been done in a backscattering geometry. For focusing the incident light on the sample an objective lens (NA = 0.9) has been used, reducing the diameter of the spot to about 0.7 μm, allowing to address individual NWs. The polarization of the scattered light has not been analyzed. The penetration depth of the applied laser light in the materials investigated here is larger than 100 nm, thus the measured Raman signal is collected from the whole interior of the NW. Simultaneously, contributions from both the (In,Ga)As core and the (Ga,Mn)As shell can be observed.

Most of the examined NWs have been scanned along their axes. The most representative Raman spectrum obtained from the middle part of individual NW is shown in Fig. 6.

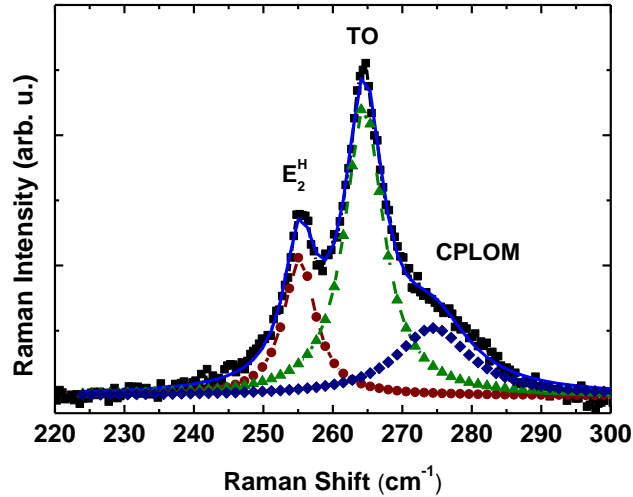

**Figure 6.** Typical high spectral resolution Raman spectrum collected for a single (In,Ga)As/(Ga,Mn)As NW using 532 nm laser light (squares) with individual Lorentzian contributions deconvoluted. The sum of the three contributions is marked as a solid blue line. The two most prominent features: GaAs-like TO phonon at 264.5 cm$^{-1}$ (triangles) and $E_2^H$ mode at 255.0 cm$^{-1}$ (dots) are the fingerprint of the wurtzite structure of the material. The higher energy shoulder signalizes sizable hole density in (Ga,Mn)As shell (diamonds).

The Raman spectrum is composed of three peaks at 255.0, 264.5 and 274.5 cm$^{-1}$. Minor variations of the values given above along nearly the entire length of each wire, or from wire to wire are observed, the latter do not exceed 1 cm$^{-1}$. On the other hand, when the laser spot is placed at the top-end part of the NW additional peaks are observed. They can be related to a rich chemical constitution of the gold enriched top part of the NW where the growth actually takes place. The detailed analysis of these results is out of the scope of this paper mainly because this part constitutes no more than the 5% of the total volume of the investigated NWs and thus can be excluded from quantitative analysis.

The feature commonly observed in the RS spectra is a weak shoulder centered at about 274.5 cm$^{-1}$. This can be interpreted as a hole-concentration-induced, coupled plasmon-LO-phonon (CPLOM) mode, resulting from the presence of (Ga,Mn)As shell.[29,30]

In the RS spectrum corresponding to (Ga,Mn)As with zinc-blende structure and Mn content close to about 1-2% the CPLOM mode is in between strongly suppressed LO and TO modes.[29] The further increase of the Mn content results in a shift of the CPLOM mode towards (or in a limit case even above) the TO frequency. As one should expect, a similar situation takes place in wurtzite (Ga,Mn)As where the only difference will be the additional $E_2^H$ mode, characteristic for this type of the crystal structure.[31] From the frequency position of the CPLOM mode it is possible to estimate the free hole concentration. The maximum of this mode in the collected RS spectrum occurs at 274.5 cm$^{-1}$. In the zinc blende (Ga,Mn)As this mode frequency is observed for the hole density of about $p = 3\times10^{19}$ cm$^{-3}$.[30] Such a value of $p$ corresponds roughly to the critical hole concentration for the metal-insulator transition in (Ga,Mn)As (see e.g. Ref. 42), and so is indicative that effects related to carriers' localization may determine the magnetic properties of these NWs, as shown further in the text.

The strongest peak observed in the RS spectrum at 264.5 cm$^{-1}$ can be related to a TO GaAs–like phonon mode, whereas the lowest energy one (at 255.0 cm$^{-1}$) to the $E_2^H$ wurtzite mode. These two phonon modes are shifted by about 2.5 and 4.0 cm$^{-1}$, respectively, compared to the predicted frequency of wurtzite GaAs.[31] Each of these modes can be attributed to the sum of contributions resulting from RS on two wurtzite NW constituents. Assuming that the principal contribution to the TO mode is related to the (In,Ga)As solid solution, the In content in the NWs core can be estimated from the shift of the TO mode. The exact dependences of the frequencies of the relevant RS features on In composition in wurtzite (In,Ga)As are yet to be determined.

Thus in order to estimate the In contents in (In,Ga)As cores the dependencies established for zinc blende material have been used.[32] According to Ref. 32 a frequency shift of the TO mode equal to 2.5 cm$^{-1}$ corresponds to the GaAs-like TO phonon mode of relaxed (In,Ga)As with the In content of 9%. However, both the (In,Ga)As core and the (Ga,Mn)As shell are not relaxed therefore the presence of the strain in the structure must be taken into account. Assuming the nominal In and Mn contents in the NW core and shell, respectively, the lattice mismatch between both components equals 0.46 %.[33] The (In,Ga)As core is subjected to a compressive strain whereas the (Ga,Mn)As shell is under a tensile strain. Since the average volumes of the core and that of the shell are comparable it is justified to assume that the absolute value of the relative lattice deformation 'in-plane' is the same for the core and the shell, and the only difference is their opposite sign.[34] However, since no phonon deformation potentials have been determined for wurtzite GaAs, we are not able to calculate the exact value of the TO phonon frequency shift induced by the strain. On the other hand, as it was shown recently by Signorello et. al.,[35] the ranges of the axial strain for WZ GaAs NWs and ZB GaAs NWs are comparable. Using the formula for the TO phonon frequency shift ($\Delta\Omega$) caused by the strain[36,37] for bulk zinc blende GaAs one finally obtains $\Delta\Omega$ = 0.81 cm$^{-1}$. Now, taking into account that the shift of the TO phonon mode resulting from the strain in wurtzite GaAs NWs is similar to that of the zinc-blende GaAs, and allowing the slight variation of this phonon mode position from wire to wire the In content in the cores equals (12 ± 4)%. This is in good agreement with the expected composition of (In,Ga)As core NWs (10% In) assessed from the calibrations performed for (In,Ga)As(100) planar layer growth, taking into account the lower In incorporation rate during MBE growth of Au-catalysed (In,Ga)As NWs reported by other groups, e.g. Ref. 28.

Here, it's worth mentioning that the InAs-like TO and LO modes do not occur in the measured spectra, even though they have been observed in (In,Ga)As solid solution (see for example Piao et al., Ref 32). For (In,Ga)As with low In content (about 10%) the InAs-like optical phonon modes have similar frequencies - about 230 cm$^{-1}$. In our case the observation of Raman scattering on LO mode in perfect crystal is forbidden due to the selection rules. The presence of TO phonon mode is allowed, but as the intensity of the InAs-like optical modes (in particular for a solid solution with a low In content[29]) is much smaller than that of GaAs-like modes, this TO mode is also not visible. It should also be mentioned, that due to their small intensities the InAs-like optical modes for (In,Ga)As solid solution with similar composition than that corresponding to investigated NWs have not been observed by Raman scattering, even for bulk crystals.[38]

It is also worth noting that LO phonon mode, present in wurtzite GaAs at 291 cm$^{-1}$ also is absent for the (In,Ga)As core in our Raman spectrum. This is consistent with the far from the resonance excitation conditions and the relevant Raman selection rules. Importantly, the later can be relaxed by the presence of numerous structural defects such as rotational twins, strain and high electric fields due to charged impurities or defects on the NW surface. The LO mode is indeed frequently observed in GaAs NWs under these conditions.[21] Therefore, the absence of the GaAs-like, (In,Ga)As LO mode in Raman spectra further confirms the outstanding quality of the cores of the investigated NWs.

Finally we have performed thorough investigations of magnetic properties of these NWs. Magnetic investigations of nanoobjects pose in general a considerable challenge due to their rather minute signal which has to be properly separated from a by far greater magnetic response of their carrier (usually the original substrate) and numerous contaminants inevitably accumulating on the specimen before the magnetic studies commence. This calls for dedicated methods and strict execution of the proper experimental code.[39] In the case of NWs reported here there are three major sources of adverse magnetic signal. The least troublesome is that of the GaAs substrate, as for the sufficiently good approximation its magnetic response is temperature independent and proportional to the applied magnetic field. Therefore, it can be reliably eradicated by a simple mathematical manipulation of the magnetometry data. The second signal, specific to samples which are grown on MBE sample holders requiring thermal anchoring by means of metallic glue, is the magnetic response of the ferromagnetic contaminants present in the glue itself (Mn-contaminated indium in our particular case). This usually overlooked part of the specimen produces a ferromagnetic-like response of strength frequently comparable to the magnitude of the investigated nanoobjects. The third signal comes from the unwanted film of (Ga,Mn)As(111)B which grew in between the NWs during the deposition of their shells. It is therefore advantageous to separate the NWs from the original substrate and transfer them onto another support of weak and known magnetic signature.

Two different methods of extracting NWs for the magnetic studies are employed here. In the first one, elaborated to preserve the spatial arrangement of the NWs, the sample surface is spin coated with PMMA [poly(methyl methacrylate), the e-beam resist] to a thickness of few tenths of micrometers (at least in the center) and subsequently slowly cooled down to cryogenic temperatures. There, the PMMA layer peels off taking all the NWs with itself. Such a NWs containing PMMA flake is then deposited onto a 0.2 mm thin rectangular piece of Si of beforehand established diamagnetic response. We hereafter call it sample A. The two major drawbacks of this method are: (i) a sizable volume of the PMMA is added to the investigated material and (ii) that PMMA also tears a part of the (Ga,Mn)As film. However, the presence of PMMA does not constitute any greater experimental challenge than the addition of the new Si support. Both substances exhibit diamagnetic properties, so their constant in *T* and linear in *H* contributions can be adequately evaluated at high temperatures and subtracted from the relevant low temperature measurements. The magnitude of the spurious signal of the fragments of the (Ga,Mn)As film has been established separately using the material left after preparation of sample B (described later) and found to constitute no more than 10% to the total signal of sample A, assuring that the dominant part of the established magnetic response comes from the investigated NWs. The additional advantage of this method lies in the fact that it facilitates orientation dependent studies.

Sample B is prepared in a more traditional approach in which the NWs are removed from the substrate in an ultrasonic ethanol bath and, after condensing, the solution is transferred to a similar piece of Si as used to support sample A. This method allows therefore the studies of the mere NWs, but can produce only a reference sample for direct magnetometry studies as the number of NWs in the specimen is vastly reduced and their spatial orientation is completely randomized. From analysis of the relevant SEM images we estimate that the original surface density of the NWs, $1.8 \pm 0.3 \times 10^9$ NWs/cm$^2$ (preserved in sample A), is reduced by a factor of ~200 in sample B, placing the expected magnetic response at the brink of detection ability. We note here that the remaining part of the material from which the NWs were ultrasonically removed contains mostly the unwanted (Ga,Mn)As(111)B film and therefore it allows the assessment of the magnitude of magnetic response of that film, and so to estimate its maximal contribution in sample A. Magnetic measurements were performed in Quantum Design MPMS XL Superconducting Quantum Interference Device (SQUID) magnetometer with the use of long Si strips to facilitate samples support in the magnetometer chamber.[39] The samples' moment *m* dependence on the magnetic field *H*, *m(H)*, is measured up to *H* = 20 kOe and as a function of temperature, *m(T)*, down to 2 K. All the data presented here have their relevant diamagnetic contributions evaluated at room temperature and subtracted accordingly.

The summary of the magnetic studies of sample A is presented in Fig. 7. In panel a) the low temperature sizable magnetic moment seen during field cooling at $H = 1$ kOe, and subsequently measured at the remanence (after quenching the field at 2 K) the thermoremanence measurement (TRM), as well as a highly nonlinear and hysteretic *m(H)* below $T_{ch} \cong 33$ K presented in panel b) are indicative of the low temperature FM coupling in the NWs. The magnetic hysteresis are of a good squarness at low temperatures, however the *m(H)* does not show a tendency to saturate even at 20 kOe. In fact, at 2 K $m(0) / m(20$ kOe$) < 0.3$ (not shown), indicating that no more than ¼ of Mn moments is FM coupled.

The true magnetic constitution of the NWs is revealed by results of temperature dependent studies presented in bottom panels of that figure. The clear maximum present on the zero field cooled (ZFC) *m(T)* is the main feature in panel c). It indicates that it is the blocking mechanism of otherwise superparamagnetic (SP) material that determines the magnetic response in weak magnetic fields and low temperatures. This in turn strongly suggests a magnetically composite (granular) constitution of the sample. The (blocked) SP characteristics and the very slow saturation at low *T* means that the FM coupling is maintained only locally in mesoscopic volumes which constitute magnetically ordered entities (supermoments) which are dispersed in otherwise paramagnetic host. The average size of these magnetic entities can be assessed from the magnitude of the temperature at which the maximum on the ZFC is seen. This maximum defines the so called (mean) blocking temperature ($T_B$) of the distribution, which is related to the (mean) volume (*V*) occupied by the supermoments through:

$$T_B = KV/25 \, k_B,$$

where $K$, the anisotropy constant in (Ga,Mn)As, ranges between 5000 to 50000 erg/cm$^3$,[40] and the factor 25 represents the typical acquisition time in SQUID magnetometry of about 100 s. From these numbers we obtain that the mean volume occupied by the supermoments corresponds to a sphere of a diameter between 15 to 30 nm. The lower bound fits perfectly to the typical NWs shell width, the larger values still can be realized here if the FM coupled volumes assume oblate shapes sprawled around the perimeter of the NWs shell.

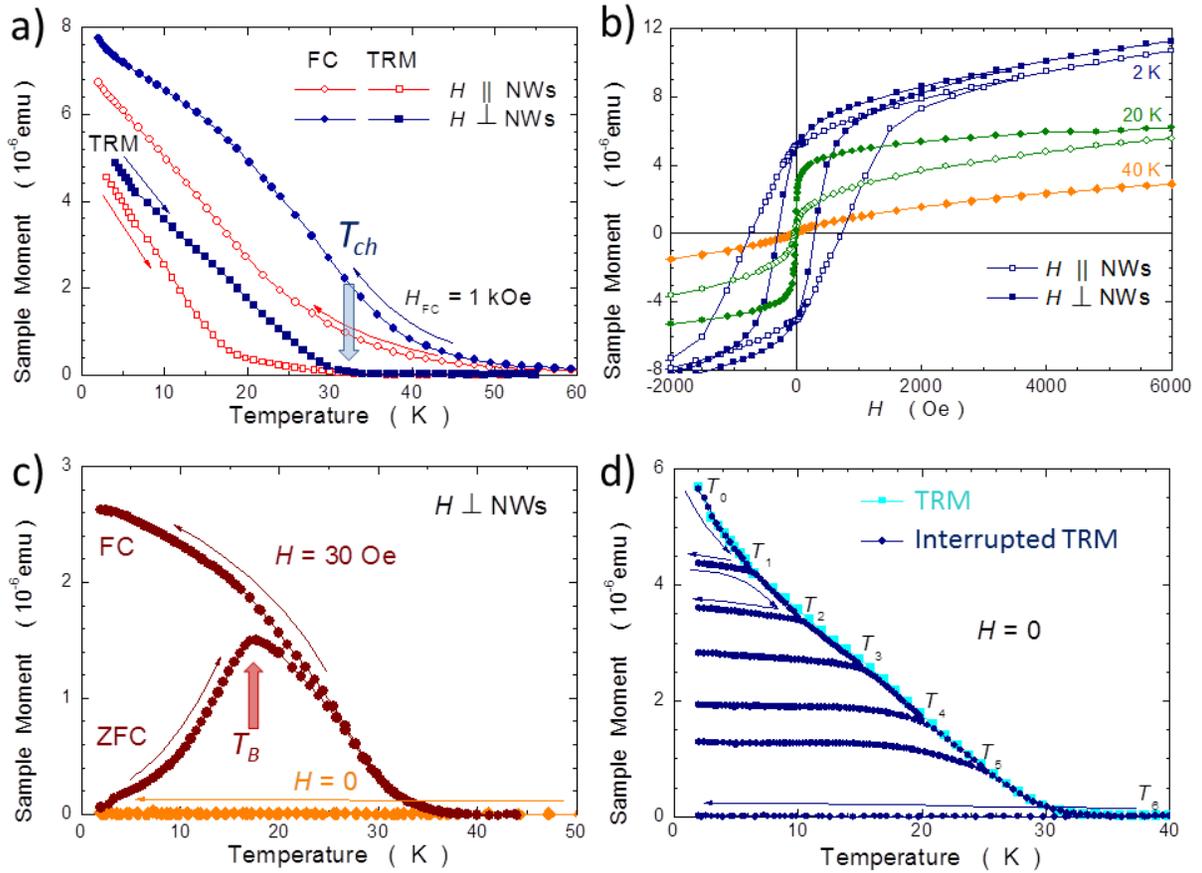

**Figure 7**. Results of magnetic studies of sample A – NWs embedded in PMMA and separated from the GaAs(111)B substrate. Open symbols: magnetic field applied along the NWs, full symbols: perpendicular to NWs. (a) Low temperature part of field cooled moment at 1 kOe (FC, circles) and at remanence on warming (TRM, squares) measured for the NWs. $T_{ch}$ marks the temperature at which the TRM moment vanishes. (b) Low field part of the magnetic hysteresis at selected temperatures. (c) Temperature dependence of zero field cooled (ZFC) and FC moment at 30 Oe (bullets) together with the magnetic moment collected during cooling at $H = 0$ (diamonds). $T_B$ marks the mean blocking temperature of this ensemble. (d) Thermal cycles of TRM consisting of a set of re-cooling from indicated intermediate temperatures ($T_i$) followed by reheating to progressively higher temperatures ($T_{i+1}$) (interrupted TRM, dark blue). This is superimposed on the typical (continuous) TRM dependence on $T$ (light blue squares).

Further evidence confirming the magnetically granular structure of the NWs comes from the lack of any moment detected from the sample at low temperatures when it is cooled from above $T_{ch}$ at $H = 0$. In both experimental configurations [for clarity panel c) shows only the in-plane case (orange diamonds)] the sample ends up at 2 K in completely demagnetized state. This is not the expected behavior for typical uniformly magnetized ferromagnetic (Ga,Mn)As, for which, even for layers at the localization boundary, on crossing $T_C$ a creation of the spontaneous moment aligned along the easy axis is commonly observed.[41] Here, the absence of this spontaneous moment demonstrates that during cooling the individual supermoments get randomly blocked behind their individual energy barriers of a height $E_B = K \cdot V$. Consistently with this scenario, after application of a small magnetic field a sizable initial increase of ZFC moment on $T$ is observed here. At the lowest temperatures only a part of the supermoments (the smallest ones) can overcome their energy barriers and get aligned by the field. As the larger ones need stronger thermal agitation, the increasing $T$ promotes the increase of $m$ until the maximum at $T_B$ is reached, above which thermal fluctuations of the supermoments reduce more $m$ than it is induced by the increase of the thermal energy. The system reaches the thermal equilibrium and becomes superparamagnetic. Importantly, below $T_B$ the blocking mechanism slows the dynamics of the supermoments what accounts for such a sizable increase of the coercivity, as witnessed in panel b).

The final evidence that the whole FM-like appearance is solely due to the slowed dynamics of the blocked supermoments is presented in panel d). Here, following the recipe presented in Fig. 3a from Ref. 41, the sample brought again to remanence at $T_0 = 2$ K is warmed up in stages through some intermediate temperatures ($T_i$), at which the sample is cooled back down to the base temperature. We note, that the sample moment gets visibly reduced only during the incremental increase of temperature (from $T_i$ to $T_{i+1}$), following the original TRM. Otherwise, it stays fairly constant, depending neither on $T$ nor on the direction it is swept. This clearly indicates the decisive role of the thermal agitation over the individual energy barriers of the supermoments on the overall magnetic response of the sample. We therefore identify $T_{ch}$ as the temperature required to reach the thermal equilibrium by all, except some statistically insignificant large, supermoments in the distribution. Importantly, this reasoning also means that the magnitude of the coupling temperature which characterizes the magnetism in the nanoclusters, their internal Curie temperature, $T_C^*$, cannot be smaller than $T_{ch}$, and so this allows us to conclude that although effective only *locally* the ferromagnetic coupling in NWs is characterized by $T_C^*$ in excess of $T_{ch} = 33$ K. However, the whole ensemble of these FM coupled volumes shows magnetic characteristics typical for blocked superparamagnets.

In view of all our characterization effort the spatially limited form of the coupling stems from the overall low hole concentration which puts the (Ga,Mn)As shell close to the metal-insulator transition, the fact already inferred from Raman scattering data. Being at the localization boundary large mesoscopic fluctuations of the local density of states are expected in the material.

Therefore a mosaic of nanometer volumes are expected, some will have increased hole density – sufficiently high to promote FM coupling, the rest will be devoid of carriers – there the Mn spins remain paramagnetic at all temperatures.[42] Importantly, the actual crystallographic structure of the material is of minor relevance for processes operating at the critical region. The presented scenario is valid for both zinc-blende or wurtzite (Ga,Mn)As and for any substrate (core) orientation. Also, the magnitude of $T_\mathrm{C}^*$ should not differ more than 10% in all these cases.[43]

The other issue is why for such a sizable Mn concentration we observe such a reduced hole concentration at the first place. In our view the most important mechanism is an enhanced incorporation of hole compensating $\mathrm{Mn_{Ga}}$ defects, as observed in planar (Ga,Mn)As(110).[44,45] The similarity of our magnetic data with other reported so far indicate that for the time being this is the price which has to be paid for high crystallinity of the (Ga,Mn)As NWs. The other effects which may contribute to the reduction of $p$ can be a low shell thickness and the occurrence of depletion layers both close to the surface and at the (In,Ga)As/(Ga,Mn)As interface region.[46,47]

The data accumulated in panel a) and b) indicate also a presence of uniaxial magnetic anisotropy with respect to the NWs axis. However, i) despite the NWs magnetize easier when the field is oriented perpendicularly to the NWs, neither for this orientation the typical easy axis behavior is seen, nor with the field aligned along the NWs the system exhibits proper hard axis characteristics; and ii) counterintuitive, the magnetic anisotropy gets weaker on lowering temperature. The magnetocrystalline anisotropy in (Ga,Mn)As depends on Mn content, concentration of valence band holes, but predominantly is governed by the epitaxial strain.[43] Importantly, due to typical substantial magnetic dilution, the shape anisotropy is not important in this family of compounds, however for sub-micrometer shapes the strain relaxation effects dominate even over the inbred magnetocrystalline anisotropy.[48] The present core-shell arrangement serves as a potentially very interesting base for the development of magnetic anisotropy. The (In,Ga)As hexagonal core defines at least 4 major crystallographic directions in the tensile strained (Ga,Mn)As shell: three perpendicular to the shell face which are oriented at 120° to one another [following {1-100} facets of the core], and the fourth one along the NW. Had the NWs been uniformly magnetized an in-shell orientation of the spins would have resulted either in a uniformly magnetized NWs along their axis (the "shape effect"), or in a spiral vortex of tangentially arranged in-shell spins. In the case investigated here according to the Zener model,[43] the tensile strained (Ga,Mn)As shell, should exert perpendicular magnetic anisotropy, and so a "cactus-like" arrangement of Mn spins could be expected.

On the other hand, the absolute values of the magnetic anisotropy observed here should depend on the details of the complex strain distribution due to anisotropic lattice mismatch in the wurtzite core/shell structure. Also, effects of faceting (and the possible presence of additional structures at the edges of the facets) may play a role here.[49] However, precise determinations of these two, and possibly other contributions, are still beyond the resolution of the available magnetic probes.

Nevertheless all our experimental findings preclude existence of the uniform ferromagnetism in the (Ga,Mn)As shells. Therefore the observed anisotropic behavior must reflect the dominant easy axis orientation of the supermoments, which according to the accumulated data tends to be along the normal to the NWs side facets. Importantly, this is in an accordance with the outlined above expectations what advocates again for the presence of ferromagnetic high-hole-density mesoscopic pools distributed within the otherwise non-conducting (Ga,Mn)As body. In this view, the weakening of the magnetic anisotropy at low temperatures can be again associated with the existence of sizably energy barriers which, without sufficient thermal agitation, oppose the rotation of the supermoments regardless of their orientation with respect to the direction of the external field.

The relevance of the analysis presented above is further confirmed by the results obtained for the reference sample B, consisting of mere NWs transferred onto a Si support, see Fig. 8. These data are essentially the same as those obtained for sample A (cf. Fig. 7 a and b), the minor differences can easily be attributed to the averaging effect over the random orientation of the NWs in sample B. It is worth to note the minute magnetic signal exerted by sample B, however its magnitude agrees within 25% with the moment valuation obtained for the average area NWs density in this sample ($10^7$ cm$^{-2}$).

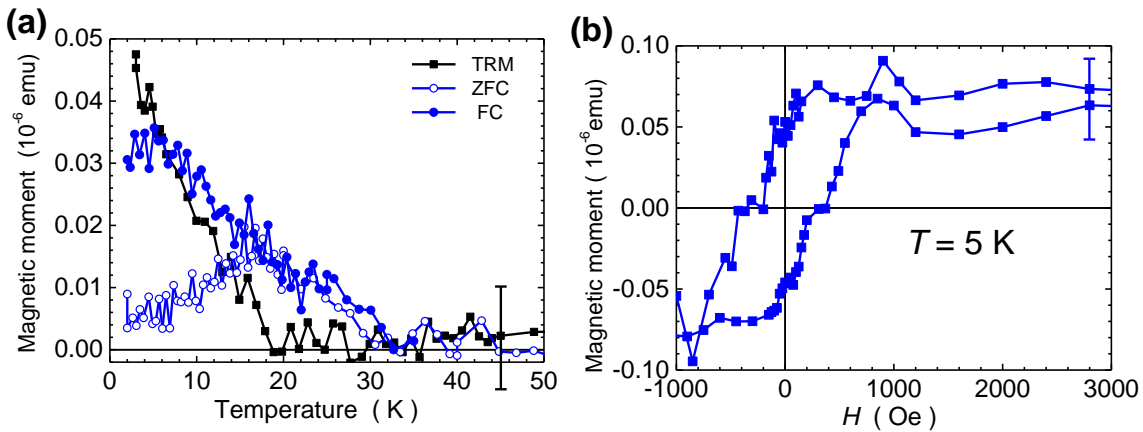

**Figure 8.** Results of magnetic studies of sample B consisting of mere NWs transferred onto a Si support. (a) - Thermoremanence (TRM, black squares), zero-field cooled (ZFC), and field cooled (FC) measurements at H = 30 Oe (open and solid blue circles, respectively). Note the same value of the mean blocking temperature, $T_B$, as seen in sample A (cf. Fig. 8 c). (b) Magnetic hysteresis loop measured at 5 K. Average error bars, as reported by the magnetometer, are indicated in each panel.

To put our finding into a broader context, we note that the data presented in panels a) and b) of Fig. 7 are qualitatively very similar to previously reported studies of other (Ga,Mn)As related NWs.[23,24,50] In these reports both the presence of a magnetic hysteresis at low temperatures and a more or less rapid roll-off of magnetic moment with increasing temperature were taken as the indication of the formation of the FM state in the NWs. One would be tempted to jump into the same conclusion here if only, as in the papers mentioned above, the results of the two upper panels of Fig. 7 were considered. However, the lower panels of this figure decisively indicate the value brought by more elaborated magnetic studies. Although we might not be entirely correct on the very nature of the magnetic coupling, the results of the for the first time performed additional examination of (Ga,Mn)As-based NWs decisively point to the dynamical effects which deceivingly produce the FM-like response, univocally precluding the transition to other thermodynamic phase both in the NWs considered here and by similarity, in the other NWs reported so far.

In conclusion we have investigated the MBE grown (In,Ga)As-(Ga,Mn)As core-shell nanowires of all-wurtzite crystalline structure. The NWs have been obtained in two steps, first (In,Ga)As NW cores have been grown at 500 $^{\circ}$C on GaAs(111)B substrate with Au catalyst; then the $Ga_{0.95}Mn_{0.05}As$ shells were deposited at low temperature (close to 220 $^{\circ}$C). The NWs are up to 1.6 um in the length and 70 nm − 100 nm in diameter. TEM microscopy investigations confirm that the (Ga,Mn)As shells are of perfect, wurtzite crystalline structure, exhibit sharp interface with (In,Ga)As cores and smooth outer surfaces. The wurtzite structure of NWs is further confirmed by Raman scattering measurements. Thorough investigations of the magnetic properties of NWs separated from the growth substrate indicate that magnetically (Ga,Mn)As NW shells are composed of scattered ferromagnetically coupled mesoscopic regions of dimensions in the range of 15 – 30 nm, which give rise to superparamagnetic properties above the blocking temperature close to 17 K and easy magnetization axis perpendicular to the NW axes. Our results point at the necessity of comprehensive measurements of the magnetic properties of NWs, preferably separated from the growth substrate. The uniform, ferromagnetic (Ga,Mn)As NWs shells can potentially be obtained if higher shell thicknesses and/or higher concentrations of substitutional Mn end up in sufficiently uniform hole concentration to promote the magnetic coupling throughout the whole body of the material.

AUTHOR INFORMATION

**Corresponding Authors**

[*]E-mail: siusys@ifpan.edu.pl , Janusz.Sadowski@maxlab.lu.se

ACKNOWLEDGMENTS

The authors acknowledge B. Kurowska and M. Bilska for preparation of the thin cross-section of NWs by FIB for TEM observations and A. Petroutchik for gold layer deposition. The MBE system used for the growth of NWs has been supported by the Swedish Research Council (VR). J.S. and M.S. acknowledge partial support by the FunDMS Advanced Grant of the ERC. A.Š. acknowledges the financial support by the SemiSpinNet EU project, European Regional Development Fund through the Innovative Economy grant (POIG.01.01.02-00-108/09), and the National Science Centre (Poland) under grant No. DEC-2011/03/B/ST3/03287. K.G and W.S. acknowledge support of the Raman scattering setup from the project DEC-2012/07/B/ST5/02080 of the National Science Center, Poland. The TEM investigations have been supported by

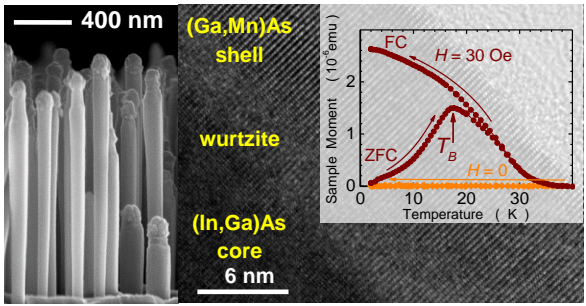

For Table of Contents Only

REFERENCES

(1) Ohno, H. *Science* **1998**, *281*, 951-956.

(2) Ohno, H.; Shen, A.; Matsukura, F.; Oiwa, A.; Endo, A.; Katsumoto, S.; Iye, Y. *Appl. Phys. Lett.* **1996**, *69*, 363-365.

(3) Slupinski, T.; Oiwa, A.; Yanagi, S.; Munekata, H. *J. Cryst. Growth* **2002**, *237*, 1326-1330

(4) Matsukura, F.; Abe, E.; Ohno, H. *J. Appl. Phys.* **2000**, *87*, 6442-6444.

(5) Csontos, M.; Mihály, G.; Jankó, B.; Wojtowicz, T.; Liu, X.; Furdyna, J. *Nat. Mater.* **2005**, *4*, 447-449.

(6) Ohno, H.; Chiba, D.; Matsukura, F.; Omiya, T.; Abe, E.; Dietl, T.; Ohno, Y.; Ohtani, K. *Nature (London)* **2000**, *408*, 944-946.

(7) Chiba, D.; Sawicki, M.; Nishitani, Y.; Nakatani, Y.; Matsukura, F.; Ohno, H. *Nature (London)* **2008**, *455*, 515-518.

(8) Stolichnov, I.; Riester, S.; Trodahl, H.; Setter, N.; Rushforth, A.; Edmonds, K.; Campion, R.; Foxon, C.; Gallagher, B.; Jungwirth, T. *Nat. Mater.* **2008**, *7*, 464-467.

(9) Chiba, D.; Ono, T.; Matsukura, F.; and Ohno, H. *Appl. Phys. Lett.* **2013**, *103*, 142418.

(10) Gryglas, M.; Kwiatkowski, A.; Baj, M.; Wasik, D.; Przybytek, J.; and Sadowski, J. *Phys. Rev. B* **2010**, *82*, 153204.

(11) Oiwa, A.; Slupinski T.; and Munekata H. *Appl. Phys. Lett.* **2001**, *78*, 518.

(12) Li, Z.; Mol, J.; Lagae, L.; Borghs, G.; Mertens, R.; Van Roy, W. *Appl. Phys. Lett.* **2008**, *92*, 112513.

(13) Rozkotová, E.; Nemec, P.; Horodyská, P.; Sprinzl, D.; Trojánek, F.; Maly, P.; Novák, V.; Olejník, K.; Cukr, M.; Jungwirth, T. *Appl. Phys. Lett.* **2008**, *92*, 122507

(14) Wunderlich, J.; Jungwirth, T.; Kaestner, B.; Irvine, A.; Shick, A.; Stone, N.; Wang, K.-Y.; Rana, U.; Giddings, A.; Foxon, C. *Phys. Rev. Lett.* **2006**, *97*, 077201.

(15) Dietl, T. *Nat. Mater.* **2010**, *9*, 965.

(16) Dietl, T.; Ohno, H. *Rev. Mod. Phys.* **2014**, *86*, 187.

(17) Yu, K.; Walukiewicz, W.; Wojtowicz, T.; Kuryliszyn, I.; Liu, X.; Sasaki, Y.; Furdyna, J. *Phys. Rev. B* **2002**, *65*, 201303.

(18) Sadowski, J.; and Domagala, J. Z. *Phys. Rev. B* **2004**, *69*, 075206.

(19) Tuomisto, F.; Pennanen, K.; Saarinen, K.; Sadowski, J. **2004**, *Phys. Rev. Lett. 93*, 055505.

(20) Parkin, S. S.; Hayashi, M.; Thomas, L. *Science* **2008**, *320*, 190.

(21) Gas, K.; Sadowski, J.; Kasama, T.; Siusys, A.; Zaleszczyk, W.; Wojciechowski, T.; Morhange, J.-F.; Altintaş, A.; Xu, H.; Szuszkiewicz, W. *Nanoscale* **2013**, *5*, 7410.

(22) Sadowski, J.; Dluzewski, P.; Kret, S.; Janik, E.; Lusakowska, E.; Kanski, J.; Presz, A.; Terki, F.; Charar, S.; Tang, D. *Nano Lett.* **2007**, *7*, 2724.

(23) Rudolph, A.; Soda, M.; Kiessling, M.; Wojtowicz, T.; Schuh, D.; Wegscheider, W.; Zweck, J.; Back, C.; Reiger, E. *Nano Lett.* **2009**, *9*, 3860.

(24) Yu, X.; Wang, H.; Pan, D.; Zhao, J.; Misuraca, J.; von Molnár, S.; Xiong, P. *Nano Lett.* **2013**, *13*, 1572.

(25) Galicka, M.; Buczko, R.; Kacman, P. *Nano Lett.* **2011**, *11*, 3319–3323

(26) Galicka, M.; Buczko, R.; Kacman, P. *The Journal of Physical Chemistry C* **2013**, *117*, 20361.

(27) Galicka, M.; Buczko, R.; Kacman, P.; Lima, E. N.; Schmidt, T. M.; Shtrikman, H. *Phys. Status Solidi RRL* **2013**, *7*, 739.

(28) Jabeen, F.; Rubini, S.; Grillo, V.; Felisari, L.; and Martelli, F. *Appl. Phys. Lett.* **2008**, *93*, 083117.

(29) Limmer, W.; Glunk, M.; Schoch, W.; Köder, A.; Kling, R.; Sauer, R.; and Waag, A. *Physica E* **2002**, *13*, 589.

(30) Seong, M.J.; Chun, S.H.; Cheong, H.M.; Samarth, N.; and Mascarenhas, A.; *Phys. Rev. B* **66**, 033202 (2002).

(31) Zardo, I.; Conesa-Boj, S.; Peiro, F.; Morante, J. R.; Arbiol, J.; Uccelli, E.; Abstreiter, G.; and Fontcuberta i Morral, A. *Phys. Rev. B* **2009**, *80*, 245324.

(32) Piao, Z. S.; Jeon, H. I.; Suh, E. –K.; Lee H. J. *Phys. Rev. B* **1994**, *50*, 18644.

(33) Sadowski, J.; Mathieu, R.; Svedlindh, P.; Domagala, J. Z.; Bak – Misiuk, J.;. Swiatek, K.; Karlsteen, M.; Kanski, J.; Ilver, L.; Åsklund, H.; Södervall, U. *Appl. Phys. Lett.* **2001**, *78*, 3271.

(34) L. Rigutti, G. Jacopin, L. Largeau, E. Galopin, A. De Luna Bugallo, F. H. Julien, J.-C. Harmand, F. Glas and M. Tchernycheva, *Phys. Rev. B* **2011**, *83*, 155320.

(35) Signorello, G.; Lörtscher, E.; Khomyakov, P.A.; Karg, S.; Dheeraj, D.L.; Gotsmann, B.; Weman H.; and Riel, H. *Nature Comm.* **2014**, *5*, 3655.

(36) Cerdeira, F.; Buchenauer, C. J.;. Pollak, F. H.; and Cardona, M. *Phys. Review B*, **1972**, 5, 580.

(37) Attolini, G.; Francesio, L.; Franzosi, P.; Pelosi, C.; Gennari, S.; and Lottici, P. P. **1994**, *J. Appl. Phys.*, *75*, 4156.

(38) Islam, M. R.; Prabhat Verma; Yamada, M.; Tatsumi, M.; Kinoshita, K.; *Proceedings of International Conference on Indium Phosphide and Related Materials, Conference*, **2001**, 129.

(39) Sawicki, M.; Stefanowicz, W.; Ney, A. *Semicond. Sci. Technol.* **2011**, *26*, 064006.(40) Sawicki, M. *J. Magn. Magn. Mater.* **2006**, 300, 1.

(41) Sawicki, M.; Chiba, D.; Korbecka, A.; Nishitani, Y.; Majewski, J. A.; Matsukura, F.; Dietl, T.; and Ohno, H.; *Nature Phys.* **2010**, *6*, 22.

(42) Dietl, T. *J. Phys. Soc. Jpn.* **2008**, *77*, 031005.

(43) Dietl, T.; Ohno, H.; Matsukura, F. *Phys. Rev. B* **2001**, *63*, 195205.

(44) Wurstbauer, U.; Sperl, M.; Soda, M.; Neumaier, D.; Schuh, D.; Bayreuther, G.; Zweck, J.; Wegscheider, W. *Appl. Phys. Lett.* **2008**, *92*, 102506.

(45) Wurstbauer, U.; Sperl, M.; Schuh, D.; Bayreuther, G.; Sadowski, J.; Wegscheider, W. *J. Crys. Growth* **2007**, *301*, 260.

(46) Proselkov, O.; Sztenkiel, D.; Stefanowicz, W.; Aleszkiewicz, M.; Sadowski, J.; Dietl, T.; Sawicki, M., *Appl. Phys. Lett.* **2012**, *100*, 262405.

(47) Storm, K.; Halvardsson, F.; Heurlin, M.; Lindgren, D.; Gustafsson, A.; Wu, P. M.; Monemar B.; and Samuelson, L. *Nature Nanotechnology* **2012**, *7*, 718

(48) Humpfner, S.; Pappert, K.; Wenisch, J.; Brunner, K.; Gould, C.; Schmidt, G.; Molenkamp, L.W; Sawicki M.; Dietl, T. *Appl. Phys. Lett.*, **2007**, *90*, 102102

(49) Spirkoska, D.; Arbiol, J.; Gustafsson, A.; Conesa-Boj, S.; Glas, F.; Zardo, I.; Heigoldt, M.; Gass, M. H.; Bleloch, A. L.; Estrade, S.; Kaniber, M.; Rossler, J.; Peiro, F.; Morante, J. R.; Abstreiter, G.; Samuelson L.; and Fontcuberta i Morral., A. *Phys. Rev B* **2009,** *80*, 245325.

(50) Bouravleuv, A.; Cirlin, G.; Sapega, V.; Werner, P.; Savin A.; and Lipsanen, H. *J. Appl. Phys.* **2013**, *113*, 144303.